\documentclass{article}

\usepackage{amsmath,amsthm,amssymb,amsfonts,latexsym,enumitem}

\usepackage{natbib}
\setcitestyle{open=(,close=)}

\usepackage{hyperref}

\newcommand{\bbbe}{\mathbb{E}}

\newtheorem{principle}{Principle}

\begin{document}

\title{Two ways game-theoretic probability can improve data analysis}
\date{\today}
\author{Glenn Shafer, Rutgers University\\\url{gshafer@business.rutgers.edu}, 
\url{www.glennshafer.com}}

\maketitle

\begin{abstract}

When testing a statistical hypothesis, is it legitimate to deliberate on the basis of initial data about whether and how to collect further data?  Game-theoretic probability's \textit{fundamental principle for testing by betting} says yes, provided that you are testing by betting and do not risk more capital than initially committed.  Standard statistical theory uses \textit{Cournot's principle}, which does not allow such \textit{optional continuation}.  Cournot's principle can be extended to allow optional continuation when testing is carried out by multiplying likelihood ratios, but the extension lacks the simplicity and generality of testing by betting.

Game-theoretic probability can also help us with descriptive data analysis.  To obtain a purely and honestly descriptive analysis using competing probability distributions, we have them bet against each other using the Kelly principle.   The place of confidence intervals is then taken by a sets of distributions that do relatively well in the competition.   In the simplest implementation, these sets coincide with R.\ A.\ Fisher's likelihood intervals.

\end{abstract}

\section{Introduction}

Game-theoretic probability studies games in which announcements by Player I define rates at which Player II can bet on outcomes. The games can have one or many rounds.  Player I's announcements can take various forms. One possibility is that Player I announces a probability distribution and Player II is authorized to buy any payoff for its expected value.

A probability distribution for a discrete-time stochastic process can be used as a strategy for Player I; it tells Player I what probability distribution for the $n$th outcome to announce after seeing the first $n-1$ outcomes.  When Player I is required to follow a particular strategy of this form, strategies for Player II define martingales.  Theorems about discrete-time stochastic processes become theorems in game theory \citep{Shafer/Vovk:2001,Shafer/Vovk:2019}.

The betting intuition brought to the fore by game-theoretic probability has proven productive for statistical theory, sometimes helping mathematical statisticians find more efficient or more powerful methods \citep{Ramdas/etal:2023}. More importantly, game-theoretic probability can help us go beyond today's standard framework for mathematical statistics, where we begin with a collection of probability distributions and test whether one of these distributions could have ``generated'' our data or try to decide which one did so.  This article discusses two ways in which game-theoretic probability goes beyond the standard framework to help make data analysis more flexible and more honest.

Considered purely as mathematics, game-theoretic probability generalizes standard probability theory, because it does not assume that any of the players are required to follow a strategy specified in advance.%
\footnote{Game-theoretic probability also allows Player I to make the more limited betting offers considered in the theory of imprecise probabilities \citep{Augustin/etal:2014,Troffaes/deCooman:2014}. I do not consider this aspect of game-theoretic probability's flexibility in this paper.}  
This allows us to authorize, in a simple and explicit way, deliberation on the basis of initial data about (1) whether and how to collect further data and (2) how to use it in statistical testing.  We already see such deliberation in practice, but squaring it with standard probability theory is not so simple and limits the tests authorized to those used in the game-theoretic picture. I discuss this point in detail in Section \ref{sec:optional}.

In Section \ref{sec:gtds}, I consider a second way game-theoretic probability can help us improve data analysis. Ever since statistical testing began to be widely used in the early 19th century, its greatest abuse has been unjustified and often clearly erroneous assumptions of randomness. Sometimes the assumption is that successive observations are chosen randomly from a population.  Sometimes the assumption is that the deviations of successive observations from a model are random -- i.e., independent of each other and of the explanatory variables in the model. In some cases, practitioners acknowledge that the assumptions are unjustified but claim that their analyses are nevertheless interesting as descriptions of data.  Because the statistical language being used is at least implicitly causal, this usually sounds like double talk. The betting picture can help here by providing a language that is clearly descriptive rather than causal.  We rank the probability distributions in a model by having them bet against each other.  This produces a set of distributions that did best in the competition.  But this set is not a ``confidence set''.  There is no suggestion that we are confident about anything, and we are certainly not confident that one of the distributions in the set ``generated'' the data. The set is simply the subset of distributions that describe the data better than the other distributions in the model.

Neither of our two improvements in data analysis provide techniques or calculations not already in use.  Just as some researchers already deliberate about how to continue experimentation, some already already use the descriptive sets we obtain in Section \ref{sec:gtds}; they are called ``likelihood intervals''.  Aside from setting aside some types of testing that cannot be legitimately authorized when we deliberate about continuation, the improvement is in how we can describe what we are doing --- how we can communicate it more clearly and more honestly to ourselves and to others.  It is also notable and important that the game-theoretic language in both cases allows us to generalize seamlessly to the case where predictions being tested or compared are not necessarily produced by probability models.  They may instead be produced by algorithms of some very different kind  --- perhaps a neural net, perhaps a physical model like those used in weather forecasting.

\section{Statistical testing with optional continuation}\label{sec:optional}

Statistical testing requires a mathematical theory of probability together with a principle that specifies how probabilities can be discredited by observations.  
\begin{itemize}
 \item
The principle used to make traditional probability theory into a theory of statistical testing is sometimes called \textit{Cournot's principle}.%
\footnote{See \citet{Shafer/Vovk:2006} and \citet{Shafer:2007,Shafer:2022} for the history of Cournot's principle.  The principle has sometimes been ridiculed by philosophers \citep{Dianconis/Skyrms:2018}.  But it has been articulated in one way or another by a panoply of mathematicians and statisticians, including Jacob Bernoulli, Antoine-Augustin Cournot himself, \'Emile Borel, Andrei Kolmgorov, Richard von Mises, and Charles Stein.}
This principle authorizes a statistician to select an event to which a probability distribution assigns small probability and to regard its happening as evidence against the distribution. 
 \item 
To make game-theoretic probability into a theory of statistical testing, we can use a principle that I have called \textit{the fundamental principle for testing by betting}.%
\footnote{Vovk and I have used various other names for this principle.  In 2001, we called it \textit{the fundamental interpretative hypothesis of probability} \citep[pp.\ 5, 14, 62]{Shafer/Vovk:2001}.  In 2019, we called it \textit{the game-theoretic version of Cournot's principle} \citep[pp.\ 226--227]{Shafer/Vovk:2019}. }
This principle, which is related to but distinct from Cournot's principle, authorizes a statistician to interpret success in betting against a probability distribution as evidence against the distribution.
\end{itemize}
Do these principles authorize optional continuation?  

As the term is used here, \textit{optional continuation} refers to the practice of deliberating, after seeing some initial data, about whether and how to continue collecting data and perhaps about how to analyze all the data together.  Such continuation may involve observations or experiments not contemplated at the outset.  It is distinguished from optional stopping, which refers only to the possibility of deciding at the outset to curtail a fully planned experiment or other study. 

The fundamental principle for testing by betting asserts the validity of optional continuation for the type of testing it considers.  Cournot's principle, in its classical formulation, does not.  It can be extended to assert the validity of optional continuation when testing is carried out by multiplying likelihood ratios, but as I will explain, the extension lacks the simplicity and generality of the fundamental principle for testing by betting.

\subsection{Optional continuation in practice and theory}

Optional continuation has long been part of statistical practice.  It is implicit, for example, in the idea of meta-analysis.  But it has proven difficult to bring it under the purview of statistical theory.

The term ``optional continuation'' with the meaning used here first appeared in print in Allard Hendriksen's master's thesis at the University of Leiden, written under the supervision of Peter Gr\"unwald \citep{Hendriksen:2017}.
Hendriksen wrote on page 3 of the thesis, 
\begin{quote}
``Optional continuation'' is the practice of combining evidence of studies that were done because of promising results of previous research on the same subject.
\end{quote}
The term has subsequently been used in other work by G\"unwald's machine-learning research group at CWI in Amsterdam \citep{Grunwald/etal:2021,Grunwald/etal:2023}.  But as of June 13, 2023, it had not yet appeared in any of the 34 statistics journals in JSTOR.

The older term ``optional stopping'' was introduced by the Duke mathematician Joseph Albert \citet{Greenwood:1938}.  Greenwood sought empirical adjustments to account for the way Joseph Rhine's laboratory was conducting and analyzing its experiments on extra-sensory perception.  Rhine stopped experimenting with each subject when a success rate thought to be statistically significant was achieved, then combined the $z$-scores achieved by successive subjects.

Greenwood's problem was brought to wider attention in mathematical statistics by William Feller's critiques of the ESP work in \citep[pp.\ 272, 286--292]{Feller:1940} and \citep[pp.\ 140, 190, 197]{Feller:1950}.  In subsequent work in probability, ``optional stopping'' has referred to stopping rules that can be adopted in advance without annulling a desired property of a stochastic process, usually the property of being a martingale \citep{Doob:1953}. 

In his book on sequential analysis \citep{Wald:1947}, Abraham Wald considered only ``sequential sampling plans'' chosen in advance.  While allowing  early stopping when there was enough evidence to make a decision, these plans specified whether or not to stop and how to continue if stopping was not mandated, all as a function of outcomes so far.  In a review of the book, George Barnard wrote that sequential analysis marked ``the entry of statistical considerations into the very process of experimentation itself'' \citep{Barnard:1947}. We know that the process of experimentation often involves not only plans adopted in advance but also opportunistic changes in plans, based on new insights and unexpected information.  

Barnard seems not to have followed up on his insight concerning the role of statistics in the process of experimentation; he does not discuss it, for example in his major article on statistical inference \citep{Barnard:1949}.  But in a subsequent article entitled ``Sequential experimentation'', R.\ A.\ Fisher wrote about the need for sequential deliberation in these terms \citep[p.\ 183]{Fisher:1952}:
\begin{quote}
The present use of the term sequential is intended to be of a broader import than the formal use of the word as associated with the systematic procedure known as sequential analysis.  The experimenter does not regard his material as wholly passive but instead looks to what may be learnt from it with a view to the improvement and extension of the enquiry. This willingness to learn from it how to proceed is the essential quality of sequential procedures. Wald introduced the sequential test, but the sequential idea is much older. For example, what is the policy of a research unit? It is that in time we may learn to do better and follow up our more promising results. The essence of sequential experimentation is a series of experiments each of which depends on what has gone before. For example, in a sample survey scheme, as explained by Yates, a pilot survey is intended to supply a basis for efficiently planning the subsequent stages of a survey. \dots
\end{quote}
Until the recent work on optional continuation, this insight about statistical practice has remained outside the ambit of statistical theory.

\subsection{A betting game with optional continuation}\label{sec:gtstat}

The simplest game used in game-theoretic probability has three players: Forecaster makes probability predictions, Skeptic bets against them, and Reality announces the outcomes.   The game is a perfect-information game, in the sense that Forecaster and Skeptic move in turn and see each other's moves.  We can vary the rules of the game, but we need not impose any further condition on what information any player might have or acquire in the course of the game, or how the players might collaborate.  Forecaster and Skeptic might be the same person.   Forecaster and Reality might be the same person.

If Forecaster keeps forecasting, Skeptic can keep betting.  Forecaster need not follow a plan or strategy about what to forecast next or how to forecast it.%
\footnote{To see how probability's limit theorems can be generalized to accommodate Forecaster's freedom, see \citet[\S7.5]{Shafer/Vovk:2019}.}
Even if Forecaster follows a strategy, Skeptic need not have a plan or strategy for when or how to bet on the forecasts.  Thus optional continuation is built into the game, for both Forecaster and Skeptic.  Skeptic can decide whether and how to continue selecting from Forecaster's betting offers, but Forecaster can decide what experiments or observations to make and what forecasts (perhaps probabilities) to give for them.  

Vovk and I have used the example of quantum mechanics to illustrate game-theoretic probability's capacity for optional continuation; see \citet[pp.\ 189--191]{Shafer/Vovk:2001} and \citet[pp.\ 215--217]{Shafer/Vovk:2019}.  In this example, we split Forecaster into two players, Observer and Quantum Mechanics.  Observer selects the experiment, and Quantum Mechanics makes the probability forecast. Formally, the game continues indefinitely, but both Observer and Skeptic can effectively stop it by making null moves.

Although optional continuation is built into the game, we need this principle to use the game in statistical testing: 
\begin{principle}[Fundamental principle for testing by betting]
Successive bets against a forecaster that begin with unit capital and never risk more discredit the forecaster to the extent that the final capital is large.%
\footnote{I first formulated the principle in this way in my SIPTA lectures \citep{Shafer:2020}.}
\end{principle}
\noindent
In one sense, this says it all.  But some elaboration may be useful:
\begin{enumerate}
\item
The principle is \textit{fundamental}, not the consequence of some more extensive philosophy or methodology.  We do not begin by saying that the forecaster's probabilities are or should be objective, subjective, personal, ``frequentist'', or whatever.  We are testing the forecaster qua forecaster, and so we are testing his forecasts qua forecasts; the question is only whether they are good forecasts, relative to the knowledge and skill of whoever is doing the testing.
\item
The forecaster may give a probability for a single event $A$, a probability distribution for an outcome $X$, or something less than a probability or a probability distribution:
\begin{itemize}
\item
If the forecaster gives a probability, you may bet on either side at the corresponding odds.
\item
If the forecaster gives a probability distribution for $X$, you may buy or sell any payoff $S(X)$ for its expected value.
\item
If the forecaster gives only an estimate $E$ of $X$, you may buy or sell $X$ for $E$.
\item
If the forecaster repeatedly gives a new probability for $A$ or new estimate for $X$, say daily, you may buy or sell tomorrow’s price for today’s price.
\item
If the forecaster gives upper and lower previsions, you may buy at the upper or sell at the lower.
\end{itemize}
\item
You \textit{begin with unit capital} only for mathematical convenience.  The discredit is measured by the ratio (final capital)/(initial capital).
\item
If you make several bets against the same forecaster (or the same theory or closely related theories), each starting with its own capital, then you are not allowed to report only the cases where you discredited the forecaster.  Instead, you must report the overall result, the sum of your final capital over all the bets divided by the sum of your initial capital over all the bets.
\item
When betting against successive forecasts, each bet uses only the capital remaining from the previous bet.  You may not borrow or otherwise raise more capital in order to continue betting.  This is what \textit{never risk more} than the initial capital means.
\item
When you stop, you must compare your initial capital with your \textit{final} capital.  You cannot claim to have discredited the forecaster because you had reached a higher level of capital in the interim. You do not have the money if you kept betting and lost it.%
\footnote{The anonymous 13th-century author who left us with the earliest surviving calculation of the chances for a throw of three dice warned us \citep[p.\ 172]{Hexter/etal:2020}:  ``Addeque, quod lusor se continuare lucrando nescit, perdendo nescit dimittere ludum.''  Not knowing how to maintain his luck when winning, the gambler does not know how to quit when losing.}
\end{enumerate}

I have stated the fundamental principle for testing by betting in 26 words,  then taken a page to explain it.  Is the principle simple?  In any case, it is coherent and teachable.  In contexts where the forecasts are only single probabilities or estimates, the principle can be taught even to those who have never studied mathematical probability.  Moreover, the principle builds on ideas about betting that most people acquire before ever studying mathematical probability.  Too many predictions contradicted by experience discredit the person making them.  If you lose too much money betting on something, you are not much of an expert about it.  Etc.

\subsection{Cournot's principle in classical form}

What principles must we add to traditional probability theory to allow optional continuation? 

Before answering this question, we answer a more basic question:  How are we authorized to discredit a probability distribution $P$ using observations?  The classical answer is Cournot's principle: we select an event $E$ that has small probability $P(E)$ (call $E$ our \textit{test event}).  The probability distribution $P$ is discredited if $E$ happens; we prefer to believe that the probabilities are incorrect rather than think that this improbable event happened.  
\begin{principle}[Cournot's principle]\label{principle:Cournot} 
If we specify an event $E$ in advance, and $E$ happens, then we may take $\alpha$, the probability of $E$, as a measure of evidence against $P$.
The magnitude of discredit is measured by how small $\alpha$ and thus how large $1/\alpha$ is.   
\end{principle}
\noindent
We may call $1/\alpha$ our \textit{test score}:
\begin{equation}\label{eq:Escore}
   \text{test score} = 
      \begin{cases}
         1/\alpha  & \text{if $E$ happen}\\
           0       &  \text{if $E$ does not happen}.
      \end{cases}
\end{equation}

Although Cournot's principle has long been fundamental to statistical theory, current philosophical fashion has made it difficult to teach.  A frequent objection is that some event of small probability always happens.  When we hear this objection, we emphasize ``specified in advance'', which requires less emphasis in game-theoretic probability, because a bet, by definition, is made in advance.  

In some cases, we may substitute ``simple to describe'' for ``specified in advance''. This is also implicit to some extent in game-theoretic probability, because a bet cannot be made and implemented unless the event is relatively simple.  

Cournot's principle can be considered a special case of the fundamental principle for testing, because $1/\alpha$ is the capital that would result from $E$'s happening if you bet unit capital on $E$.

\subsection{Extending Cournot's principle to test variables}

This extension of Cournot's principle does not require us to specify in an advance a goal $1/\alpha$ for the strength of the evidence.  

Suppose $S$ is a nonnegative random variable, chosen in advance and so not too hard to describe, with $\bbbe_P(S)=1$ (call $S$ our \textit{test variable}). Our next principle says that a realized value $s$ of $S$ discredits $P$ to the extent that $s$ is much larger than $1$.  
\begin{principle}[Authorization to test with a test variable]\label{principle:testvariable}
If we specify a test variable $S$ in advance, then we may take $s$, the observed value of $S$, as a measure of evidence against $P$.  We then interpret $s$ (our \textit{test score}) on the same scale as we use in Cournot's principle.  In other words, when $s=1/\alpha$, it has the same weight against $P$ as the happening of a pre-specified event $E$ when $P(E)=\alpha$.
\end{principle}
\noindent
Cournot's principle is the special case of Principle \ref{principle:testvariable} where $S$ is given by \eqref{eq:Escore}.  Principle \ref{principle:testvariable} adds the possibility of a more graduated report on the strength of the evidence against $P$.%
\footnote{A more widely used way of obtaining a more graduated report is to use p-values; see \S\ref{sec:pvalue}.} 

It might seem that the greater flexibility offered by a test variable $S$ comes at a price.  When $s$ is the realized value, the events $\{S=s\}$ and $\{S\le s\}$ happen, and Markov's inequality tells us that our score $1/P(E)$ would have been at least as great, often greater, had we chosen one of these events as our test event $E$.  But of course we could not have made these choices, because we did not know $s$ in advance.

Like the classical form of Cournot's principle, Principle 3 can be considered a special case of the fundamental principle for testing by betting.  The observed value $s$ of the test variable $S$ is the capital that would result from buying $S$ for its expected value.

\subsection{Extending Cournot's principle to test martingales}

Now suppose we want to test a probability distribution $P$ for a stochastic process $X:=X_1,X_2,\dots$, and we observe the $X_t$ successively.  We use a \textit{test martingale}, a nonnegative martingale $S_1,S_2,\dots$ with $\bbbe_P(S_1)=1$, again chosen in advance and hence relatively simple.  The value $s_t$ of $S_t$ may become known to us only when we have observed $X_1,\dots,X_t$.  To interpret $s_t$, we adopt this principle:
\begin{principle}[Authorization to test with a test martingale]\label{principle:testmartingale}
If we specify a test martingale $S_1,S_2,\dots$ in advance, then at all times $t$ we may take $s_t$, the observed value of $S_t$, as the current measure of evidence against $P$.  If we want, we may stop at time $t$ and continue to regard $s_t$ as our measure of evidence against $P$.  We may interpret each $s_t$ (each \textit{test score}) on the same scale as we use in Principles \ref{principle:Cournot} and \ref{principle:testvariable}.  In other words, when $s_t=1/\alpha$, it has the same weight against $P$ as the happening of a pre-specified event $E$ with $P(E)=\alpha$.
\end{principle}
\noindent
Principle \ref{principle:testvariable} is the special case of Principle \ref{principle:testmartingale} where $P$ says that all the $S_t$ are equal to each other, so that nothing can be accomplished by continuing past $t=1$.

Like Cournot's principle and our previous extensions of it, Principle \ref{principle:testmartingale} can be considered a special case of the fundamental principle of testing by betting.  For each $t$, $s_t$ is the capital obtained at time $t$ if we first buy $S_1$ and then at every step invest all our winnings so far in $S_t$.  But we are still testing a mathematical object, a probability distribution $P$.  Forecaster is following a fixed strategy, which tells him to use $P$'s successive conditional probabilities as forecasts, and Skeptic's strategy (the test martingale) is also specified in advance. Neither Forecaster nor Skeptic has a role that allows them to exercise optional continuation.  So Principle \ref{principle:testmartingale} is not a principle of optional continuation in the sense of this article.  It is, however, a principle of optional stopping.

Although the statement of \ref{principle:testmartingale} does not mention betting, I do not recall seeing the principle explained or advocated without a betting story.

\subsection{Improvised testing (optional continuation for Skeptic)}

Principle \ref{principle:testmartingale} authorizes the statistician to use a  test martingale specified in advance.  Improvisation is not yet authorized.  For this, we need some further principle.  As with Principle \ref{principle:testmartingale}, we are testing a probability distribution $P$ for a stochastic process $X:=X_1,X_2,\dots$, and we observe the $X_t$ successively.  When $x_1,\dots,x_{t-1}$ are possible values of $X_1,\dots,X_{t-1}$, we call a nonnegative variable $S(X_t)$ a \textit{round-$t$ test variable given $x_1,\dots,x_{t-1}$} if $\bbbe_P(S(X_t)|x_1,\dots,x_{t-1})=1$; when $t=1$, this reduces to $\bbbe_P(S(X_1))=1$.  We can formulate a principle for improvisation in testing as follows:
\begin{principle}[Authorization to wing it when testing]\label{principle:improvisetest}
Suppose we set $s_0=1$, specify a round-1 test variable, say $S_1(X_1)$, and then, beginning with $t=1$,
\begin{enumerate}
\item
we observe $X_t$'s value $x_t$,
\item
we set $s_t:=s_{t-1}S_t(x_t)$, and
\item
we specify a round-$(t+1)$ test variable given $x_1,\dots,x_t$, say $S_{t+1}(X_{t+1})$. 
\end{enumerate}
Suppose we continue for as long as we want and stop whenever we want (after step 2 for some $t$).  Then at all times $t$ until after we stop, we may take $s_t$ as the current measure of evidence against $P$.  We may interpret $s_t$ on the same scale as we use in Principles \ref{principle:Cournot}, \ref{principle:testvariable}, and \ref{principle:testmartingale}.
\end{principle}

Principle \ref{principle:improvisetest} generalizes Principle \ref{principle:testmartingale}, and like Principle \ref{principle:testmartingale}, it can be considered a special case of the fundamental principle for testing by betting.  Skeptic is now a free player, not constrained to follow a strategy specified in advance.

\subsection{Improvised probabilities (optional continuation for Forecaster)}

Principle \ref{principle:improvisetest} authorizes a statistician testing a probability distribution to improvise.  But this still does not bring us to R.\ A. Fisher's vision, where the statistician helps construct over time not only a test but also the probabilities being tested.  In this vision, statistician and scientists brainstorm to design an experiment with outcome $X_1$, to which they assign probabilities based on some theory they want to test, and after observing $X_1=x_1$, they brainstorm again about what they have learned and design a possibly unanticipated experiment with outcome $X_2$, and so on.

It is tempting to try to square traditional probability with Barnard's vision by imagining that this collaboration defines a probability distribution $P$ progressively.  The first design includes a probability distribution $P_1$ for $X_1$.  The second includes a probability distribution $P_2$ for $X_2$, etc.  The product $P_1\times \cdots \times P_k$, where $k$ is where the research team stops, is a probability distribution $P$.  

But the statistician did not set out to test $P_1\times \cdots \times P_k$.  She and her colleagues waited to design the second experiment and its 
$X_2$ and $P_2$ until they had seen $x_1$.  Had $x_1$ come out differently, their subsequent brainstorming might have produced a different $X_2$ and $P_2$, and so on.  If there is a probability distribution being tested,  it would seem to involve conditional probabilities for $X_2$ given all the different $x_1$ that might be observed (and perhaps also all the other ways the research team's information and thinking might evolve while the first experiment was being performed).  And so on.

Some decades ago  A. Philip \citep{Dawid:1984,Dawid:1991} bravely argued that these dependencies should not matter---that we can design significance tests, confidence intervals, and Bayesian procedures that are unaffected by probabilities, somehow true or somehow invented, involving the might-have-beens.  As these might-have-beens do not matter, we can just pretend that we have the requisite independence.  This is Dawid's \textit{prequential} model.  Although some statisticians (including myself) found it appealing, others found it confusing.  What are we really testing?  Are we testing a huge and not fully specified probability distribution $P$ whose unspecified probabilities include probabilities for actions of the research team doing the testing?

Leaving all this aside, can we formulate a principle that authorizes us to use Dawid's insight to construct test scores?  Here's a try.

\begin{principle}[A prequential testing principle]\label{principle:prequential}
Suppose we set $s_0=1$, construct an experiment that will produce a variable $X_1$, a probability distribution $P_1$ for $X_1$, and a test variable $S_1$ for $P_1$, and then, beginning with $t=1$,
\begin{enumerate}
\item
we observe $X_t$'s value $x_t$,
\item
we set $s_t:=s_{t-1}S_t(x_t)$, and
\item
we construct an experiment (perhaps newly conceived) that will yield a variable $X_{t+1}$, a probability distribution $P_{t+1}$ for $X_{t+1}$, and a test variable $S_{t+1}$ for $P_{t+1}$. 
\end{enumerate}
Suppose we continue for as long as we want and stop whenever we want (after step 2 for some $t$).  Then at all times $t$ until after we stop, we may take $s_t$ as the current measure of evidence against the $P_t$ we have constructed so far all being valid.  We may interpret $s_t$ on the same scale as we use in Principles \ref{principle:Cournot}, \ref{principle:testvariable}, \ref{principle:testmartingale}, and \ref{principle:improvisetest}.
\end{principle}

Principle \ref{principle:improvisetest} is a special case of Principle \ref{principle:prequential}.  And Principle \ref{principle:prequential}, like our preceding extensions of Cournot's principle, can be considered a special case of the fundamental principle for testing by betting.  Now both Forecaster and Skeptic are free agents, not constrained to follow any strategy specified in advance.

The principle's consistency with testing in the game-theoretic framework is not surprising, as that framework was partly inspired by Dawid's prequential model.

\subsection{The role of Ville's inequality.}\label{sec:pvalue}

Ville's inequality says that if $S_1,S_2,\dots$ is a test martingale, then 
\[
        P\left(\sup_{t\ge1} S_t\ge \frac{1}{\alpha}\right)\le \alpha.
\]
Some people (including myself) have sometimes said that Ville's inequality authorizes optional continuation.  This is a careless formulation.  First because a theorem is never more than mathematics; it cannot authorize anything.  Secondly because the principle it suggests is not an optional continuation principle developed in this article.

Ville's inequality tells us that $1/\sup_{t\ge1} S_t$ is a ``p-variable'' and so $1/\sup_{t\ge1} s_t$ is a p-value.  Well, almost.  It is at least implicit in the notion of a p-value, as statisticians understand and use the term, that we have observed it and know we have observed it.  We do not expect this for $1/\sup_{t\ge1} s_t$.  But we do observe upper bounds.  At time $t$, we have observed the upper bound $1/\sup_{1 \le i \le t} s_i$, and an upper bound on a p-value is a p-value.  So most statisticians who use p-values would probably accept this principle:
\begin{principle}[The dynamic p-value principle]\label{principle:pvalue}
As we continue to make observations, we may always use the current $1/\sup_{1 \le i \le t} s_i$ just as statisticians usually use a p-value.
\end{principle}

\noindent
This principle is implicit in the use of confidence sequences, which go back to \citet{Darling/Robbins:1967b}.

Principle \ref{principle:pvalue} can be considered an optional stopping principle, because it authorizes us to use $1/\sup_{1 \le i \le t} s_i$ like a p-value if we stop at time $t$.  But it is more than an optional stopping principle, because it authorizes us both to use $1/\sup_{1 \le i \le t} s_i$ like a p-value at time $t$ \textit{and also} to continue.  It is not an optional continuation principle in the sense of this article, because it does not authorize us to change the later experiment (the probabilities for future $X$s) or the test martingale.

As an optional stopping principle, Principle \ref{principle:pvalue} can be compared with Principle \ref{principle:testmartingale}. Neither is stronger than the other.  Principle \ref{principle:testmartingale} authorizes us to use $s_t$ as a measure of our evidence against $P$ and to continue doing so if we stop.  But it does not allow us to continue using $s_t$ if we do not stop and hence does not authorize us to use the sometimes larger $\sup_{1 \le i \le t} s_i$ \citep{Shafer/etal:2011}.  But it gives $1/s_t$ the force of a fixed significance level, which is greater than the force of a p-value.

Ville's inequality and Principle \ref{principle:pvalue} have generalizations in game-theoretic statistics, where they use game-theoretic definitions of upper and lower probability and expected value \citep[Exercise 2.10]{Shafer/Vovk:2019}.

\subsection{Conclusion}

We have shown that the traditional principle for testing a probability distribution (Cournot's principle) can be extended so that it fully accommodates optional continuation and yet does not explicitly use game-theoretic probability or ideas about betting.  Is this extension worth the trouble?

The clear message of the exercise is that the fundamental principle of testing by betting, coupled with game-theoretic probability, provides a theoretical basis for optional continuation that is simpler, clearer, and more general.  Readers will judge for themselves, but I submit that Principle \ref{principle:prequential} is overly complex, ill-motivated, and impossible to teach without reference to betting. It remains, moreover, less general than the fundamental principle of testing by betting, because it requires Forecaster's moves on each round of a forecasting game to be a probability distribution.

Our exercise has also illustrated the new clarity brought to statistical theory by game-theoretic probability's distinction between Forecaster and Skeptic.  This distinction has helped us see the complexity of the notion of optional continuation.  Optional continuation for Forecaster is a step further than optional continuation for Skeptic.

\section{Game-theoretic descriptive data analysis}\label{sec:gtds}

Researchers often construct statistical models that cannot be taken seriously as anything more than approximate descriptions of their \emph{study populations} --- the populations for which they have data.  Unfortunately, our methodology and terminology for constructing such models (estimation, significance tests, confidence intervals, credible regions, etc.) can only be understood in terms of inferences about larger populations or observations not yet made.

A study population is often merely a \emph{convenience sample} (examples we managed to find).  Sometimes it is an \emph{entire population} (perhaps the eight highly industrialized nations, or the five hundred corporations in the S\&P 500 index).  Describing such populations means summarizing --- identifying general features rather than details.  When a study population is merely a collection of numbers, we may be able to summarize it by giving a few statistics, such as the average $\overline{y}$ and the standard deviation $s$. 

The difficulty I am discussing arises as soon as we ask about the precision of descriptive statistics.  If $\overline{y}=5.346$, then is $5$ just as good a description?  What about $6$ or $4$?  Maybe.  If the numbers are very spread out, then saying that $0$ is in the middle might be just as good as saying that $5.346$ is in the middle. Our usual response to this difficulty is to replace $\overline{y}$ with a confidence interval, such as $\overline{y}\pm 1.96 s/\sqrt{n}$, but then we are pretending to make an inference to a theoretical mean or a larger population.

The first thesis of this section is that when our goal is merely to describe a study population, we should use methods of forecasting rather than methods of inference. An average of a collection of numbers is a forecast of each number, and asking whether other summaries forecast  as well or nearly as well as the exact average does not involve inference to some larger population.

A second thesis is that game-theoretic probability can be used to develop this notion of internal forecasting into a general methodology for descriptive statistics. In this general methodology, the forecasts are probabilities or probability distributions, and the relative success of these forecasts is evaluated by having them bet against each other.

\subsection{Theory}

As explained in Section \ref{sec:optional}, game-theoretic probability uses probability distributions as forecasting strategies. It recasts the notion of independent observations as a property a forecasting strategy might have: it always makes the same forecast or uses the same forecasting rule. The question whether given observations are random with respect to a probability distribution is replaced by the question whether a forecasting strategy withstands bets against its forecasts.

In addition to serving as a forecasting strategy, a probability distribution can also serve as a strategy for betting against a probability forecaster \citep[Ch.\ 10]{Shafer/Vovk:2019}, \citep{Shafer:2021}.  This use of a probability distribution is called \textit{Kelly betting}.  In the simplest case, it leads to interpreting a likelihood ratio as the payoff of a gamble.  

This duality in way probability distributions can be used --- as forecasting strategies and as strategies for betting against forecasts, is key to the simple idea I am proposing here:  Choose a statistical model (i.e., a collection of probability distributions), have them bet against each other's forecasts for the study population, and take the distribution or distributions that do best in the competition as our description of the study population.  The result is purely descriptive; it does not suppose that the winners will forecast well in other data, and it does not even make any claims about the model we have chosen being best for the study population.  The choice may be purely conventional.

\paragraph{From probability forecasts to description.}  

Any method of description requires relatively arbitrary choices and conventions. Here we begin by choosing variables that have values for each individual in our study population:  a variable $Y$ (the \emph{target} variable) and variables $X_1,\dots,X_K$ (the \emph{forecasting} variables).  Inferential statistical theory has accustomed us to think of these choices as causal modeling, but when description is our goal, $Y$ and the $X_k$ are simply what we want to describe.  For whatever reason, we want to know how they vary together in the study population.

Next we choose a family of algorithms $(P_\theta)_{\theta\in\Theta}$, where each algorithm $P_\theta$ uses the $X_k$ to give probability distributions for $Y$.  The $P_\theta$ are our forecasters, and the probability distributions they give are our forecasts. 

We will want to limit the complexity of the family $(P_\theta)_{\theta\in\Theta}$.  In inferential statistics, simplicity is said to be a virtue of a model because complex models are likely to overfit --- i.e., not to generalize to other study populations.  When are goal is description rather than inference, either because we are uninterested in other populations or because we cannot make assumptions that would justify inference to other populations, a more immediate virtue of simplicity is salient.  Description is description only when it is simple enough to be understood.

Being familiar and conventional is another virtue of a descriptive forecasting family.  Description requires convention, and it can only be communicated to those who know the convention.

Once the variables and the forecasting family are chosen, we can evaluate the $\theta$ according to their relative forecasting success within the study population.  Let $\Theta_{\text{D}}$ be the subset of $\Theta$ consisting of $\theta$ whose forecasts perform reasonably well in our judgment; we may call it the \emph{description range}.  Its elements are the \emph{descriptive forecasters of $Y$}.  Often we will be most interested in some particular aspect of the descriptive forecasters.  When $\theta=(\mu,\sigma^2)$, for example, we might be interested in $\mu$.  In other cases, we might be interested in the difference $Y_\text{A}-Y_\text{B}$ when values of the $X_k$ for two individuals A and B are given.  In general, the range of $h(\theta)$ when $\theta$ ranges over $\Theta_{\text{D}}$ is our \emph{description range} for $h(\theta)$.

\paragraph{The game-theoretic competition.}

For each ordered pair $(\theta_1,\theta_2$ of elements of $\Theta$, we pit $\theta_2$ bet against $\theta_1$.  We do this by imagining that Forecaster has the distribution $P_{\theta_1}$ and Skeptic has the distribution $P_{\theta_2}$.  Skeptic has unit capital.  Forecaster offers to sell Skeptic any non-negative payoff $S(Y)$ for $E_{\theta_1}(S)$.  In deciding what payoff $S$ to buy, Skeptic can use his distribution $P_{\theta_2}$ in different ways, but here I focus on the simplest way:  He buys the payoff $S$ given by 
\begin{equation}\label{eq:Kelly}
    S(Y):=P_{\theta_2}(Y)/P_{\theta_1}(Y)
\end{equation}
at the price Forecaster requires:
\[
         E_{\theta_1}\left(\frac{P_{\theta_2}(Y)}{P_{\theta_1}(Y)}\right)=1.
\]
This bet turns Skeptic's initial unit capital into $S(y)=P_{\theta_2}(y)/P_{\theta_1}(y)$.  So this ratio provides a measure of $\theta_2$'s forecasting success relative to $\theta_1$.

The choice \eqref{eq:Kelly} for $S$ is an example of \emph{Kelly betting}.  It has well known optimization properties for Skeptic when Skeptic is confident in $Q$ as a forecast.%
\footnote{See \citet{Breiman:1961}.  For more on Kelly betting, see \citet[Chapter 10]{Ethier:2010} and \citet{Ziemba:2015}.  For other roles Kelly betting can play in statistical theory, see \citet{Shafer:2021} and \citet{Vovk/Wang:2021}.}  
An important alternative to Kelly betting is fractional Kelly betting, which risks only a fraction of one's capital.  Being more cautious, this penalizes Skeptic less when $P_{\theta_2}(y)/P_{\theta_1}(y)$ is low.  In the sequential version of our theory, where the individuals in the study population are forecast sequentially, this would make the competition between forecasters less sensitive to individuals that are unusual with respect to the whole study population.%
\footnote{The sequential version is emphasized in most work on discrete-time game-theoretic probability; it is more interesting because it allows forecaster $\theta$'s forecast for the next individual to depend other information from inside the game (previous individuals' values of $Y$), on information from outside the game, or even on whim.  We leave all this aside here because it does not give different results when the forecasting family is specified and we use Kelly betting.}

The ratio $P_{\theta_2}(y)/P_{\theta_1}(y)$ is familiar to statisticians under the name \emph{likelihood ratio}.  We will usually chose our forecasting family so that a unique maximum of $P_\theta(y)$ exists.  The value of $\theta$ that achieves the maximum, say $\hat{\theta}$, is the \emph{maximum-likelihood estimate}, and the likelihood ratio 
\[
      L(\theta) = \frac{P_\theta(y)}{P_{\hat{\theta}}(y)}
\]
is a number between zero and one that assesses the forecasting performance of the other $\theta$.  

R.\ A.\ Fisher's use of the word \emph{likelihood} has endured for a century, and no mathematical statistician will be able to put the word out of their mind while reading the rest of this paper.  Because Fisher gave the word an inferential rather than descriptive meaning, I will make a point of avoiding it as much as possible.  But I will use the familiar notation: $\hat{\theta}$, $L(\theta)$, and also $l(\theta)$ for $\ln(L(\theta))$.

A number of authors, including A.\ W.\ F.\ \citet{Edwards:1972} and Richard \citet{Royall:1997}, have adopted Fisher's proposal to use intervals of $L(\theta)$ as inferences. These authors sometimes call $L(\theta)$ a measure of \emph{support} for hypotheses about $\theta$.  Some of the computational work by these authors may be useful for implementing the proposal advanced here.  But I will avoid the word \emph{support} just as I avoid the word \emph{likelihood}.

\paragraph{Fisher's cutoffs.}

Using cutoffs suggested by Fisher in 1956,%
\footnote{See \citet[p.\ 71]{Fisher:1956} or page 75 of the posthumous third edition (1973).  The names suggested here (good, fair, poor, acceptable) are not Fisher's.  He did say that a value less than $1/15$th was ``open to grave suspicion''.}
we may classify the performance of the $\theta$ according to their value of $L(\theta)$ as follows:
\begin{center}
\begin{tabular}{lc}
Good &
$L(\theta)\ge 1/2$\\
Fair &
$1/2 > L(\theta)\ge 1/5$\\
Poor &
$1/5 > L(\theta)\ge 1/15$\\
Unacceptable &
$1/15 > L(\theta)$
\end{tabular}
\end{center}
These cutoffs are arbitrary, but no more so than the $5\%$ and $1\%$ frequencies used for statistical significance.  If equally accepted as conventions, they can be equally serviceable.  Their meaning in terms of betting will be readily understood by the public, especially those not already trained in mathematical statistics.

\subsection{Examples}\label{sec:examples}

I offer three very simple examples.  The first is purely formal and about as simple as possible:  the forecast is a single probability, always the same probability.  The second involves a classic convenience sample.  The third, a fictional instantiation of a problem of current interest, involves an entire population.

It would be very useful to study examples with more complex data and correspondingly complex models, such as stochastic processes and regression, and to have computational resources for dealing with such examples.   I hope this paper will inspire work in that direction.

\paragraph{Forecasting with a single probability.}

Suppose we observe successive trials of an event, and each algorithm in our forecasting family has a fixed probability that it uses each time as its forecast.  Formally, $\Theta=(0,1)$, and Forecaster $\theta$ always gives $\theta$ as its forecast. 

If we observe $100$ trials, and the event happens $70$ times, then $\hat{\theta}=0.7$, and 
\[
    L(\theta) := \left(\frac{\theta}{0.7}\right)^{70}
                      \left(\frac{1-\theta}{0.3}\right)^{30}.
\]
Our scheme for rating the forecasters yields these approximate description ranges:
\begin{center}
\begin{tabular}{lc}
Good &
$0.64 < \theta < 0.76$\\
Fair or better &
$0.61 < \theta < 0.78$\\
Acceptable &
$0.59 <\theta  < 0.80$
\end{tabular}
\end{center}
The forecaster $\theta=1/2$ may have been of particular interest, and we may want to emphasize that its performance was unacceptable.

Not surprisingly, Fisher's categories are roughly consistent with inferential practice.  The standard error of the maximum-likelihood estimate $0.7$ is $0.046$, suggesting a $95\%$ confidence interval of $(0.61,0.79)$, significantly different from $1/2$.  But unlike this confidence interval, the description ranges merely describe.  The forecasting family does not say that the trials of the event are in any sense independent.  The description ranges tell us merely which constant forecasts perform relatively well in the data.

\paragraph{Fourier's masculine generation.}

The calculation of error probabilities from statistical data was first made possible by Laplace's central limit theorem, and the calculation was first explained to statisticians by Joseph Fourier (1768--1830).
 
Fourier had been an impassioned participant in the French revolution and an administrator under Napoleon.  After the royalists regained power, a former student rescued him from impoverishment with an appointment to the Paris statistics bureau.  This assignment left him time to perfect the theory of heat diffusion for which he is best known, but as part of his work at the statistics bureau, he published two marvelously clear essays on the use of probability in statistics, in 1826 and 1829. According to Bernard Bru, Marie-France Bru, and Olivier Bienaym\'e, these were the only works on mathematical probability read by statisticians in the early 19th century.%
\footnote{See \citet[p.~198]{Bru/etal:1997}. The former student was Chabrol de Volvic.  The annual reports issued by the bureau during Fourier's tenure list no editor on their cover pages.  Fourier was no doubt primarily responsible for editing them, and I identify him as the editor in the references \cite{Fourier:1826,Fourier:1829}.}

To illustrate Laplace's asymptotic theory, Fourier studied data on births and marriages gleaned from 18th-century parish and governmental records in Paris.  He was particularly interested in the length of a masculine generation --- the average time, for fathers of sons, from the father's birth to the birth of his first son.  On the basis of $505$ cases, he estimated this average time to be $33.31$ years.  In his bureau's report for 1829, he gave the bounds on the estimate's error shown in Table~\ref{ta:error}.%
\footnote{Table 64 of the bureau's report for 1829 \citep[pp.\ 143ff]{Fourier:1829}.}

\begin{table}
\caption{Fourier's error probabilities.  The first two lines give the probabilities for errors \emph{in months} that Fourier calculated for his estimate of $33.31$ \emph{years}.  He gave $1/20$, for example, as the probability of the estimate erring by more than $11.4516$ months. The second two lines put this information into the language of confidence intervals.  The 95\% interval, for example, is 32.6 to 34.0 years.}\label{ta:error}
\vspace*{-3mm}
\[
\begin{matrix}
 \text{probability} & 1/2            &         1/20   &      1/200     &     1/2000     &    1/20000\\
 \text{error}         & \pm 2.7528 & \pm7.9932 & \pm11.4516 & \pm14.2044 & \pm16.5480\\
 &&&&&\\   
 \text{level}          & 50\%         & 95\%          &     99.5\%    &     99.95\%  &      99.995\%\\
 \text{interval}      & (33.1,33.5) & (32.6,34.0) &  (32.4,34.3)  & (32.1,34.5)  & (31.9,34.7)
\end{matrix}
\]
\hrule
\end{table}

Laplace's theory is applicable, of course, only if the 505 cases are mutually independent, and they are not.  They are a convenience sample, not a random sample from all 18th-century fathers of sons in Paris.  This convenience sample is of interest, but Fourier's inferential analysis of it is unjustified.  A descriptive analysis is needed.  

For description, we do not need Fourier's implicit assumption that the 505 cases were a random sample.   We need a conventional forecasting family.  Let us use the most conventional family, the normal family with mean $\mu$ and variance $\sigma^2$.  Here $\theta=(\mu,\sigma^2)$ and
$
                         \hat{\theta}= (\overline{y},s) = (33.31,7.642),
$
where $\overline{y}$ is the average of the 505 ages, and $s$ is their standard deviation.  From the information Fourier gives us, we know that $\overline{y}=33.31$ years and $s=7.642$ years.%
\footnote{Fourier did not report the data, but we can calculate the standard deviation from the error probabilities he gave.}

Writing $l(\theta)$ for $\ln(L(\theta))$, we have 
\begin{equation}\label{eq:log}
  l(\mu,\sigma^2) = N\left(\ln(s) - \ln(\sigma)
                       - \frac{s^2+(\overline{y}-\mu)^2}{2\sigma^2} + \frac12\right),
\end{equation}
where $N=505$.  The values of $(\mu,\sigma^2)$ for which \eqref{eq:log} exceeds $\ln(1/2)$ constitute the good description range for $(\mu,\sigma^2)$, those for which it exceeds $\ln(1/5)$ the fair or better range, those for which it exceeds $\ln(1/15)$  the acceptable range.

Following Fourier, we are interested only in description ranges for $\mu$.  So our question is what values of $\mu$ are included in these three description ranges.  So we maximize \eqref{eq:log} for each $\mu$, by setting $\sigma^2$ equal to $s^2+(\overline{y}-\mu)^2$.%
\footnote{In inferential likelihood theory, the result of maximizing a likelihood over an unwanted parameter is sometimes called a ``profile likelihood'' \citep[p.\ 158]{Royall:1997}.} 
This gives
\begin{equation}\label{eq:relfourier}
    l(\mu,s^2+(\overline{y}-\mu)^2) = N\left(\ln(s) - \frac12 \ln(s^2+(\overline{y}-\mu)^2)\right).
\end{equation}
This is greater than $\ln(1/C)$ when $\mu$ is in the interval
\begin{equation}\label{eq:foreint}
    \overline{y} \pm s\sqrt{\left(2C\right)^\frac1N - 1}.
\end{equation}
Table~\ref{ta:four2} uses \eqref{eq:foreint} to calculate description ranges.

\begin{table}
\caption{Description ranges for the masculine generation in Fourier's study population, calculated from \eqref{eq:foreint}.}\label{ta:four2}
\begin{center}
\vspace*{-2mm}
\renewcommand*{\arraystretch}{1.2}
\begin{tabular}{clc}
   $C$      &     & description range\\
  $2$  &  Good  &           $(32.9,33.7)$  \\
  $5$  &  Fair or better  &  $(32.8,33.8)$  \\
  $15$ &  Acceptable  &  $(32.7,33.9)$  
\end{tabular}
\hrule
\end{center}
\end{table}

Comparing Tables \ref{ta:error} and \ref{ta:four2}, we see that Fourier's 95\% confidence interval includes values that are not quite acceptable according to this descriptive analysis:  $(32.6,34.0)$ is a little wider than $(32.7,33.9)$.


\paragraph{A fictional survey of perceptions.}

Some organizations in the United States have recently surveyed their employees about perceptions of discrimination.  To avoid the complexities involved in real examples, consider the following fictional example.  

An organization wants to know whether its employees of different genders and racial identities differ systematically in their perception of discrimination.  Most of the employees respond to a survey asking whether they have suffered discrimination because of their gender or race.  The employees saying yes are distributed as shown in Table~\ref{ta:four1}.

According to the usual test for the difference between two proportions, the difference between the rows (male vs female) and the difference between the columns (BIPOC vs White) are both statistically significant at the 5\% level.  But the $20$ percentage-point difference between BIPOC males and BIPOC females is not, as its standard error is 
\[
     \sqrt{\frac13 \frac23\left(\frac{1}{20}+\frac{1}{10}\right)}\approx 0.18 = 18 \text{ percentage points}.
\]
These simple significance tests seem informative.  The differences declared statistically significant seem general enough to be regarded as features of the organization, but we hesitate to say this about the difference declared not statistically significant.

Yet the theory of significance testing does not fit the occasion.  Have the individuals in the study (or their responses to the survey) been chosen at random from some larger population?  Certainly not.  For anyone who has been inside an organization long enough to see its employees come and go, seeing or guessing the reasons, the idea that they constitute a random sample is phantasmagoria.  Nor can we agree that their responses are independent with respect to some ``data-generating mechanism''.  Many of them see the same media and talk with each other.

If we took the theory of significance testing seriously for this example, we would also worry about multiple testing.  The 5\% error rate we claim for our tests is valid under the theory's assumptions only when we make just a single comparison.  We have made three comparisons and might make more.  

The theory's assumptions are not met, and we have abused the theory.  But there is a larger issue.  The theory is irrelevant from the outset, because its goals are irrelevant.  The organization did not undertake the survey in order to make inferences about a larger or a different population or about some data-generating mechanism.  The organization wanted only to know about itself.  It wanted to know how its employees' perceptions vary with gender and race.  This calls for a descriptive analysis.

\begin{table}\caption{Numbers and proportions of positive responses, in a fictional study of the employees of a fictional organization, to the question whether one has experienced discrimination in the organization as the result of one's identity.  Here BIPOC means Black, indigenous, and people of color.}\label{ta:four1}
\vspace*{-4mm}
\begingroup
\renewcommand*{\arraystretch}{2.4}
\[\arraycolsep=10pt
\begin{matrix}
  & \text{Female} & \text{Male}  &  \text{Totals}\\
  \text{BIPOC} & 
  \displaystyle{\frac{8}{10}}=80\%   & 
  \displaystyle{\frac{12}{20}} =60\%   & 
  \displaystyle{\frac{20}{30}}\approx67\%\\
  \text{White}  & 
  \displaystyle{\frac{20}{50}}=40\% & 
  \displaystyle{\frac{20}{120}} \approx17\% & 
  \displaystyle{\frac{40}{170}}\approx24\%\\
  \text{Totals} & 
  \displaystyle{\frac{28}{60}}\approx47\% & 
  \displaystyle{\frac{32}{140}} \approx23\% & 
  \displaystyle{\frac{60}{200}}=30\% 
\end{matrix}
\]
\endgroup
\hrule
\end{table}

For a descriptive analysis, we can use the obvious forecasting family, in which each cell in the $2\times2$ table has its own forecast:
\[
   \theta = (\theta_\text{bf},\theta_\text{bm},\theta_\text{wf},\theta_\text{wm}),
\]
where $\theta_\text{bf}$ is the forecast that a BIPOC female will say yes to the survey, etc.  According to the data in Table~\ref{ta:four1}, 
\[
    \hat{\theta} = \left(\frac{8}{10},\frac{12}{20},\frac{20}{50},\frac{20}{120}\right),
\]
and 
\begin{align*}
    L(\theta) = &\left(\frac{\theta_\text{bf}}{4/5}\right)^8
                      \left(\frac{(1-\theta_\text{bf})}{1/5}\right)^2
                     \left(\frac{\theta_\text{bm}}{3/5}\right)^{12}
                     \left(\frac{(1-\theta_\text{bm})}{2/5}\right)^8\\
                     &\left(\frac{\theta_\text{wf}}{2/5}\right)^{20}
                     \left(\frac{(1-\theta_\text{wf})}{3/5}\right)^{30}
                     \left(\frac{\theta_\text{wm}}{1/6}\right)^{20}
                     \left(\frac{(1-\theta_\text{wm})}{5/6}\right)^{100}.
\end{align*}

We found earlier that the $20$ percentage-point difference between BIPOC males and BIPOC females is not statistically significant.  In this descriptive analysis, the question can be reframed this way:  what differences between BIPOC males and BIPOC females within the study population are forecast by good forecasters?  We can answer the question by looking at all the $\theta=(\theta_\text{bf},\theta_\text{bm},\theta_\text{wf},\theta_\text{wm})$ that rank as good forecasters by having a value of $L(\theta)$ greater than $1/2$ and finding the range of their values for $\theta_\text{bf}-\theta_\text{bm}$.  The range is from a little more than $0$ to about $0.4$.  We can say that there are good forecasters who give nearly the same forecasts for the two groups.

When the individuals responding to a yes-no survey are categorized in more than one way, or when other data is collected about them, we may prefer to use a more sophisticated forecasting family, such as logistic regression.  The logic will remain the same.  For particular interesting values of the forecasting variables, we can calculate the range of forecasts given by good forecasters.  We can similarly calculate ranges for odds ratios.  The computations involved are not trivial, but software environments adequate to the task would not be need to be more complex for the user than those that now use logistic regression for nominally inferential analyses.

The descriptive approach can be compared with the inferential approach used in 2016 by the University of Michigan's Diversity, Equity \& Inclusion Initiative.   Michigan sought inferential legitimacy by using random samples.  As they explained in their report on the faculty survey \citep[p.\ 6]{Michigan:2017},
\begin{quote}
Given the large faculty population at the University of Michigan, this study used a sample survey approach rather than a census of all faculty. A carefully selected sample, with randomization, allows researchers to make scientifically based inferences to the population as a whole.
\end{quote}
The second sentence of this quotation raises the question, unanswered by the authors, of how they would have made scientifically based inferences had they performed a whole census.  How would they then have decided which differences were meaningful?

In any case, the authors chose $1{,}500$ out of $6{,}700$ faculty members at random to complete the survey.  The survey results were then analyzed using logistic regression, and a number of differences were observed to be statistically significant.  It was found, for example, that female faculty were 130\% more likely to feel discriminated against than male faculty (i.e., the odds ratio for a positive response to the question was equal to 2.3 and significantly different from 1).  The results of the survey were clearly meaningful, but the inferential logic is problematic.  As David A.\ Freedman has shown, randomization probabilities do not justify logistic regression \citep{Freedman:2008}. Our descriptive theory is not affected by this problem and is just as applicable to a complete census as to a random or non-random sample.

\subsection{Discussion}

There is no need to document here the persistence and prevalence over the past two centuries of the use of inferential methods with non-random samples.  These abuses have been repeatedly documented and deplored.%
\footnote{David Freedman's survey, \citep[pp.\ 212--217]{Freedman:2009}, is particularly concise and revealing. See also \citet{Alexander:2015,Berk/Freedman:2003,Bru/etal:1997,Freedman:1991,Mason:1991,Matthews:1995,Shafer:2019}.}
It seems appropriate, however, to acknowledge other efforts to find descriptive alternatives.

One obvious alternative is to fit models but simply to refrain from significance tests and confidence intervals.  This has often been the practice in fields, such as geodesy, that have been more influenced by the Gaussian tradition than by the Laplacean tradition.%
\footnote{Marie-Fran\c{c}oise Jozeau has documented the competition of the two traditions in geodesy \citep{Jozeau:1997,Shafer:2019b}.}  
A surveyor's customers, for example, need a boundary line, not a confidence band.  Efforts to establish the same practice in the social sciences have usually had limited and ephemeral impact.  During much of the 20th century, some sociologists avoided using significance testing except for random samples, but this avoidance did not endure.   In a study of the use of significance testing in two prominent sociology journals from 1935 to 2000 \citep{Leahey:2005},
Erin Leahey found a shift around 1975 among authors who had data on an entire population.  Before that date, some of these authors declined to use significance tests, afterwards few did.  

John Tukey's advocacy of data analysis was another effort to shift attention from inference to description \citep{Donoho:2017}.
But the distinction between exploratory data analysis and confirmatory data analysis that emerged from this effort kept description in a subsidiary role.  Once you calculate confidence intervals or significance tests, you are suggesting that goal is a confirmatory analysis.  Mere description is no longer the goal.

In 2004, in his \textit{Regression Analysis: A Constructive Critique} \citep{Berk:2004}, the sociologist Richard A.\ Berk gave three cheers for a purely descriptive use of multiple regression:  estimate the regression coefficients, but do not calculate p-values or confidence intervals.  The book has been widely cited, but so far as I know few researchers have followed this advice.  What use is a regression coefficient if you have no sense of its precision?

In 1995, Berk and his colleagues Bruce Western and Robert E.\ Weiss proposed that the neo-Bayesian philosophy could justify statistical modeling for entire populations \citep{Berk/Western/Weiss:1995}. I have not been able to understand their argument.

Another notable effort to make inferential tools descriptive was mounted by David A.\ Freedman and David A.\ Lane in 1983 \citep{Freedman/Lane:1983}. In the simplest cases, their proposal involves permuting residuals and interpreting a p-value as a measure of how unusual the actual data is in population of alternative data thus generated.  Along with many other mathematical statisticians, I was very intrigued by this proposal when it appeared, but I always found its logic elusive.

The most common justification of using inferential tools with non-random samples is, of course, the argument that assumptions are never exactly satisfied.   George Box's slogan, ``all models are wrong but some are useful'' is evoked to justify calculating p-values and confidence intervals.  This leaves unanswered, however, the question of what we are being asked to have confidence in.  Any claim that the calculations are mere description is contradicted by the language being used.

\section{Acknowledgments}

I have had helpful discussions on these topics with Marshall Abrams, John Aldrich, Gert de Cooman, Harry Crane, Nancy DiTomaso, Ruobin Gong, Peter Gr\"unwald, Zev Hirsch, Wouter Koolen, Kitae Kum, Barry Loewer, Aaditya Ramdas, Judith ter Schure, Vladimir Vovk, Ruodu Wang, Sandy Zabell, and Snow Zhang.  On the topic of optional continuation, the workshop \textit{Safe, Anytime-Valid Inference (SAVI) and Game-theoretic Statistics} (May 30--June 3, 2022 in Eindhoven, Netherlands) was especially useful.

\addcontentsline{toc}{section}{References}

\bibliographystyle{plainnat}
\bibliography{nejsds}

\end{document}